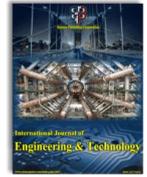

# Computer science students simulation in capturing tacit knowledge by using NGT for reducing traffic jam

**Leon Andretti Abdillah[1]\*, Ifit Novita Sari[2], Dian Eka Indriani[3]**

*[1]Computer Science Faculty, Universitas Bina Darma, Palembang, Indonesia*
*[2]Program Pascasarjana S2 Pendidikan Ilmu Pengetahuan Sosial, Universitas Kanjuruhan Malang, Indonesia*
*[3]Pancasila and Civic Culture Education, STKIP PGRI Bangkalan, Bangkalan 69116, Indonesia*
*\*Corresponding author E-mail:leon.abdillah@yahoo.com*

**Abstract**

The subject knowledge management systems is one of the main courses in information systems study program, faculty of computer science. This course is offered at A concentration for students in semester 5 (five). Knowledge consists of knowledge explicit and tacit. To capture tacit knowledge, it can be done by involving a number of methods, namely: 1) Brainstorming, and 2) Nominal Group Technique. The study also involves a number of social information technologies, such as: 1) Facebook, 2) WordPress, 3) DropBox, and 4) YouTube. The topic for the theme of knowledge is how to reduce traffic jam. After passing 2 (two) simulation rounds, this research get 8 (eight) ideas suggestion related to vehicle parking in an effort to reduce traffic jam. The tacit knowledge capture simulation with NGT is able to provide acceptable suggestions by all the panelists involved. The use of social information technology in this study also received a very good response from the panelists.

## 1. Introduction

One of the most recent studies in the field of information systems (IS) is related to knowledge management systems (KMS). The progress of an organization depends on how the organization can manage its knowledge optimally in accordance with the needs of the organization and the development of the times. The term knowledge management system is a synonym for organizational memory system [1]. Many knowledge management initiatives rely on information technology as an important enabler [2].

Information and Communication Technologies (ICT) may play an important role in effectuating the knowledge-based view of the firm by enhancing the firm's capability to manage the knowledge it possesses [3]. KMSs have become the systems that maintain corporate history, experience and expertise that long-term employees hold [4]. Institutions that survive in long term is an institution that has the ability to manage the company's knowledge by using the latest leading technology.

In general, knowledge is divided into 2 (two) groups, namely: 1) Explicit Knowledge, and 2) Tacit Knowledge [5]. Explicit knowledge is knowledge already available as: a) Form, b) Document, c) Report, d) Book, e) Recipe, f) Guide, g) Manual, or h) Code Program. While Tacit Knowledge is a knowledge that can not be used directly. Tacit knowledge exists in people's heads and is extremely difficult to transfer [6]. One characteristic of tacit knowledge is that it is personal knowledge [7]. Tacit Knowledge is still in the form of: a) Ideas, b) Experience, c) Intelligence, d) Intuition, or d) Thought. knowledge management exists to transform [8] tacit knowledge into knowledge that is easily communicated and documented (explicit knowledge).

One of the most crutial phase in managing KMS is caputring "Tacit Knowledge". Knowledge capturing techniques can capture knowledge when it flows between experts [9] who are concurrently engaged in a selected project's phase. This phase can be done using the method: 1) Brainstroming, and 2) Nominal Group Technique (NGT).

Successfulgroupbrainstorming involves aniterative processes [10] that consist of 2 (two) main stages: 1) idea exchange at the sociallevel, and 2) idea expansion at the cognitivelevel. In this article, barinstorming will be included as the first stage among 2 (two) stage of NGT.

The NGT was developed in 1971 [11] to present a group process model for situations related to planning group that can use for: 1) identifiying strategic problems, and 2) developning appropriate and innovative progams to solve them. NGT involves four key [12] stages: 1) silent generation, 2) round robin, 3) clarification, and 4) voting (ranking).

Some previous researches have been conducted in the field of KMS, such as: 1) Managing KMS in higher education [13], 2) KMS in Assets Reconciliation [14], 3) Knowledge Sharing Infrastructure Evaluation [15].

Economic growth and the population has triggered a shift in population concentration from village to urban. Cities are currently home to nearly half of the world's population [16]. In January 2018 [17], world population are about 7.476 Billion, 54% are urbanization (table 1). Metropolitan cities such as Jakarta, Surabaya, Medan, Makassar, Palembang, or Tangerang have been populated by more than 1 (one) million inhabitants. Even Jakarta entered into one of the most populous cities in the world [18].

**Table 1:**Digital around the World in 2018

| Indicator | Population (Billion) | Penetration |
|---|---|---|
| Internet Users | 4.021 | 53% |
| Active Social Media Users | 3.196 | 42% |
| Unique Mobile Users | 5.135 | 68% |

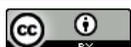

| Active Mobile Social Users | 2.958 | 39% |
|---|---|---|
| Total Population | 7.593 | 55% |

At the moment, Indonesia is a member of the G-20 [19], a country with one-third of ASEAN's GDP and almost 40 percent of its population [20]. In 2016 World Bank [21] listed Indonesia in rank number (eight) on gross domestic product ranking table based on purchasing power parity (PPP).

**Table 2:** Gross Domestic Product 2016, PPP

| Ranking | Country | Millions |
|---|---|---|
| 1 | China | 21,417,150 |
| 2 | United States | 18,569,100 |
| 3 | India | 8,702,900 |
| 4 | Japan | 5,266,444 |
| 5 | Germany | 4,028,362 |
| 6 | Russian Federation | 3,397,368 |
| 7 | Brazil | 3,141,333 |
| 8 | Indonesia | 3,032,090 |
| 9 | United Kingdom | 2,796,732 |
| 10 | France | 2,773,932 |

Purchasing power parity also includes the ability to buy motor vehicles, both two-wheel and four-wheeled. In the region of ASEAN, about 3.1 million vehicles were sold in 2016. Indonesia is the biggest consumer with about one million cars sold in 2016 [22].

The recent increase in cars is the main cause of traffic congestion [23]. The growth of motor vehicle rate which is not proportional to the increase of road race has triggered the increasingly severe congestion. This bottleneck condition will be raised in this article. The rest of this article consist of 3 (three) more sections. Section 2 (two) presents methods that are used in this research. Findings or results followed by discussions are present in section 3 (three). The last section 4 (four), conclusions, covers summarization of the research.

## 2. Materials and methods

### 2.1. Research subjects

The research method applied in this research is applied research method in the teaching class. The researcher will perform the simulation of the practice of some methods that exists in a particular course topic.

The subject of the study is knowledge management systems (KMS) taught in semester 5 (five). The field of KMS studies is offered in line with the demands of the times in the field of information systems.

Object of research involved are students of study program of information systems, faculty of computer science. Those students are taking course of knowledge management systems. They are fifth graders who are taking concentration courses. The total number of students involved was 112 students, but only 100 students who involved in the simulation.

KMS methods involved in the research, namely: 1) Brainstorming, and 2) Nominal Group Technique (NGT). It consists of two rounds in which panelists rate, discuss, and then re-rating a series of items or questions [24].

### 2.2. Nominal group technique procedure

The preparation for NGT simulation in class simulation include 4 (four) aspects, as follow: 1) A class room that large enough to accommodate 4 (four) to 8 (eight) groups or around 40 (fourty) students. Organize the students to sit nearby their group member, 2) Each group needs to write down their ideas, after brainstorming session, on a blank white paper with pencil or pen, 3) Lecturer as facilitator provide a laptop to tabulate participants ideas, and display it to the class with a projector, and 4) Lecturer clarifies participants roles and group objectives, explains a statement of the importance of the task, and NGT procedures.

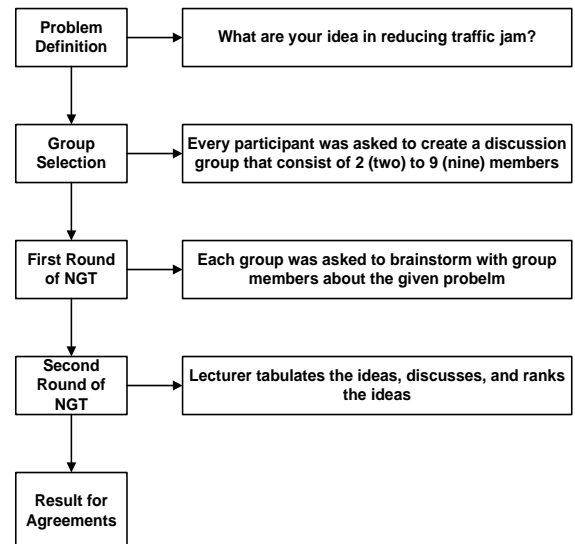

**Fig. 1:** Nominal Group Technique Flow.

### 2.3. Social information technology

During the lectures, researchers also used a number of the latest social information technologies, such as: 1) Facebook as virtual learning hub between lecturer and students, 2) WordPress is used by students to document the progress of their teamwork, 3) Drop-Box is used to store students' group reports, and 4) YouTube is used to enrich college students' course materials.

## 3. Results and discussions

### 3.1. Panelist distribution based on group and gender

Among 112 students who enrolled in the KMS class, there are 100 students participated in the class simulation (male = 54%; female = 46%). Those students will discuss a such of famous topic in urban problem "How to reducing traffic jam".

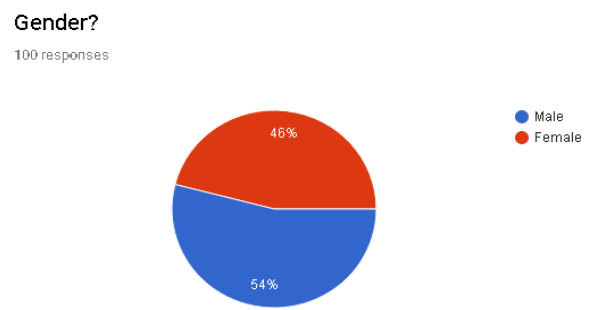

**Fig. 2:** Pannellist Gender.

Total panelists consist of 3 (three) classes and 18 groups. Author identifies the class as Alpha, Beta, and Gamma. In Alpha class consist of 6 (six) groups, Beta class consist of 8 (eight) groups, and Gamma class consist of 4 (four) groups.

**Table 3:** Group Class

| No | Class | Total Group |
|---|---|---|
| A | Alpha | 6 |
| B | Beta | 8 |
| C | Gamma | 4 |
|  | Total | 18 |

### 3.2. Brainstorming results

In the first phase, all groups are asked to do brainstorming for discussing a given particular condition. The increase in economy and population has affected the increase in the number of vehicle ownership, both wheel and motorcycles. The theme of the discussion raised this time is related to the parking of vehicles in order to reduce the traffic jam in big city. Every group will suggest 5 (five) ideas in written paper. The first round could be seen in table 4.

**Table 4:** Group Ideas

| Code | Ideas | X | Y | Z |
|---|---|---|---|---|
| A | Parking Space/Lot | 4 | 6 | 3 |
| B | Parking Security | 4 | 1 | 3 |
| C | Parking Facilities | 3 | 0 | 1 |
| D | Parking Rates | 4 | 2 | 0 |
| E | Parking Promotion | 2 | 0 | 2 |
| F | Parking Officer | 3 | 0 | 0 |
| G | Parking Systems | 1 | 0 | 3 |
| H | Parking Rules | 1 | 0 | 4 |
| I | Mass Transportation | 0 | 4 | 0 |
| J | Two Wheel Vehicles | 0 | 1 | 0 |
| K | Pedestrians Sidewalk | 0 | 1 | 0 |
| L | Bicycle | 0 | 3 | 0 |
| M | Increase Vehicles Tax | 0 | 3 | 0 |
| N | Limit Vehicle Ownership | 0 | 3 | 0 |
| O | Even-Odd Vehicle Plates | 0 | 1 | 0 |
| P | FlyOver | 0 | 1 | 0 |
| Q | Vehicle Expiration Age | 0 | 2 | 0 |
| R | Purchase Limit of BBM | 0 | 1 | 0 |
| S | Car Special Flow | 0 | 1 | 0 |
| T | Parking Location | 0 | 0 | 3 |
| U | Parking Fine | 0 | 0 | 1 |
| V | Public Road Line | 0 | 0 | 1 |
| W | Parking Point Guidance | 0 | 0 | 1 |
| X | Parking Space/Lot | 4 | 6 | 3 |

According to table 4, there are 24 (twenty four) total initial ideas in reducing traffic jam. Class-Alpha (X) accounts for a total of 8 (eight) ideas (A, B, C, D, E, F, G, H). Class-Betha (Y) accounts for a total of 14 (fourteen) ideas (A, B, D, I, J, K, L, M, N, O, P, Q, R, S, T).

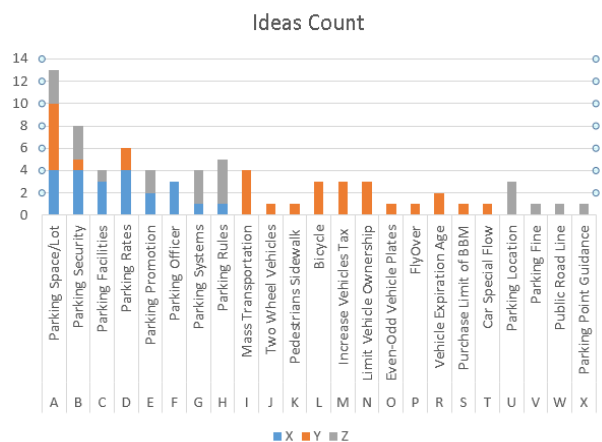

**Fig. 3:** Total Ideas Count.

Class-Gamma (Z) accounts for a total of 10 (ten) ideas (A, B, C, E, G, H, U, V, W, X). The Illustration for comprehensive group ideas count could be seen in figure 3.

### 3.3. Nominal group technique

After all of group participants submitted their ideas, then lecturer tabulated those ideas for all classes. When total count for any ideas has been tabulated, then lecturer re-ranked the ideas based on the highest count in total. After the results obtained tabulation in re-ranking, then got some ideas to overcome congestion. After all the ideas collected in the rankings are ascending, it can be seen which ideas are most widely proposed by panelists. Table 5 shows final ranking for all submitted ideas.

**Table 5:** Selected Group Ideas

| Code | Ideas | Count | Selected |
|---|---|---|---|
| A | Parking Space/Lot | 13 | * |
| B | Parking Security | 8 | * |
| D | Parking Rates | 6 | * |
| H | Parking Rules | 5 | * |
| C | Parking Facilities | 4 | * |
| E | Parking Promotion | 4 | * |
| G | Parking Systems | 4 | * |
| I | Mass Transportation | 4 | * |

From 24 (twenty four) ideas that exist, not all of them will be taken as a result of the capture of knowledge. The ideas that are proposed are ideas that have high points only. Based on new rank in table 5, the rank then ordered as follow: 1) Parking Space/Lot, 2) Parking Security, 3) Parking Rates, 4) Parking Rules, 5) Parking Facilities, 6) Parking Promotion, 7) Parking Systems, and 8) Mass Transportation. Count of each concept could be seen in table 5. After discussing with the panelists, it is agreed that 8 (eight) high-level concepts will be selected. Those selected ideas are coded with the asterisk symbol (*).

### 3.4. Evaluation of social technology uses

This research provides learners with various forms of media to connect with their instructors [25]. Those media are distributed through social information technologies, such as: 1) Facebook, 2) WordPress, 3) DropBox, and 4) YouTube.

The first evaluation is related to using Facebook as a learning environment based on social media technology. Dominant students panelist like (see figure 4) to learn by involving the application that grow significantly and attract online users [25].

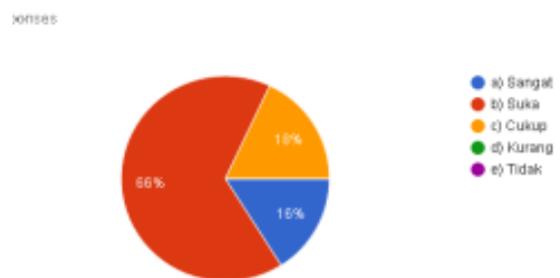

**Fig. 4:** Evaluation Result of Using Facebook.

The second evaluation is related to using Blog as a hub for students in displaying their team work. More than 65% students panelist like (see figure 5) to do their team work in fun environment where they have media to express their work, academic repositories and connections [26].

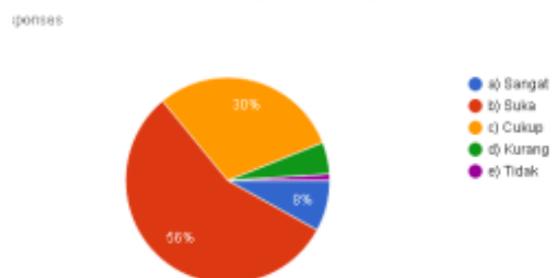

**Fig. 5:** Evaluation Result of Using Blog.

The third evaluation is related to using YouTube as a third resource in enriching their corse materials. More than 53% students panelist like (see figure 6) to enrich their course material by involving a very powerful medium in exploring lecture materials [27], [28].

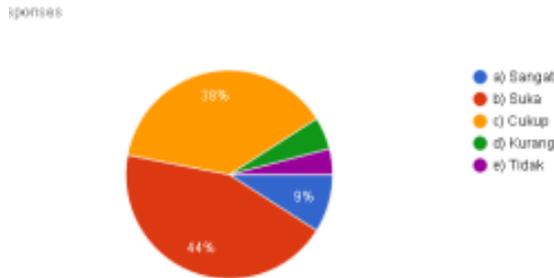

**Fig. 6:** Evaluation Result of Using Youtube.

The fourth evaluation is related to the use of DropBox as cloud repository. More than 63% students panelist like to store their team work by using DropBox (see figure 7). DropBox provides free cloud repository space for everyone to store their documents in various of formats [29]. Each submission in DropBox will has URL to make it easier in sharing. Dropbox also very handy tool [30] for its users.

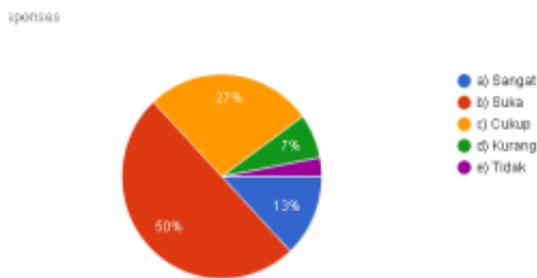

**Fig. 7:** Evaluation Result of Using Dropbox.

## 4. Conclusion

After the simulation are succeed conducted in the class, author able to summarized some point of views as follow:

Brainstorming is an excellent tool for discussing a problem. In this study was able to capture as many as 24 (twenty four) initial ideas from the all of participants. Brainstorming facilitates students corporation among their group members in class discussions.

The NGT method can be implemented well, and is able to select 8 (eight) best proposed ideas in reducing traffic congestion. Those ideas include: a) Parking Space/Lot, b) Parking Security, c) Parking Rates, d) Parking Rules, e) Parking Facilities, f) Parking Promotion, g) Parking Systems, and h) Mass Transportation. Most of the ideas are related to parking vehicles.

The involvement of popular social information technology in the lectures gained a very good response from the learners. For next research, Author interests to involve more knowledge capturing and codification methods with some tools that free available over the internet.

For next study, researcher interested to involve cloud medium and smart devices for higher education purposes.

## References

[1] R. Maier, Knowledge management systems: Information and communication technologies for knowledge management, Third ed. Heidelberg: Springer Science & Business Media, 2007.

[2] K. C. Desouza, "Barriers to effective use of knowledge management systems in software engineering," *Communications of the ACM,* vol. 46, pp. 99-101, 2003.https://doi.org/10.1145/602421.602458.

[3] H. Benbya, *et al.*, "Corporate portal: a tool for knowledge management synchronization," *International journal of information management,* vol. 24, pp. 201-220, 2004.https://doi.org/10.1016/j.ijinfomgt.2003.12.012.

[4] T.-C. Lin and C.-C. Huang, "Understanding knowledge management system usage antecedents: An integration of social cognitive theory and task technology fit," *Information & Management,* vol. 45, pp. 410-417, 2008.

[5] L. A. Abdillah, "Knowledge Management Systems," in *Computer Science for Education*, ed. Palembang: Bina Darma University, 2014.

[6] Y. M. C. Yeh, "The implementation of knowledge management system in Taiwan's higher education," *Journal of College Teaching & Learning (TLC),* vol. 2, 2011.

[7] V. Ambrosini and C. Bowman, "Tacit knowledge: Some suggestions for operationalization," *Journal of Management studies,* vol. 38, pp. 811-829, 2001.https://doi.org/10.1111/1467-6486.00260.

[8] J. Joeliaty and A. P. Aryani, "Perancangan Knowledge Management System pada Bagian Diklat PT Dirgantara Indonesia," *Jurnal Manajemen dan Bisnis Indonesia,* vol. 1, pp. 197-213, 2017.

[9] Z. Pourzolfaghar, *et al.*, "A technique to capture multi-disciplinary tacit knowledge during the conceptual design phase of a building project," *Journal of Information & Knowledge Management,* vol. 13, p. 1450013, 2014.https://doi.org/10.1142/S0219649214500130.

[10] H.-C. Wang, *et al.*, "Idea expander: supporting group brainstorming with conversationally triggered visual thinking stimuli," in *Proceedings of the 2010 ACM conference on Computer supported cooperative work*, 2010, pp. 103-106.https://doi.org/10.1145/1718918.1718938.

[11] L. Delbecq and A. H. Van de Ven, "A group process model for problem identification and program planning," *The Journal of Applied Behavioral Science,* vol. 7, pp. 466-492, 1971.https://doi.org/10.1177/002188637100700404.

[12] S. S. McMillan, *et al.*, "How to use the nominal group and Delphi techniques," *International journal of clinical pharmacy,* vol. 38, pp. 655-662, 2016.https://doi.org/10.1007/s11096-016-0257-x.

[13] L. A. Abdillah, "Managing information and knowledge sharing cultures in higher educations institutions," in *The 11th International Research Conference on Quality, Innovation, and Knowledge Management (QIK2014)*, The Trans Luxury Hotel, Bandung, Indonesia, 2014.

[14] M. Fitriyani, *et al.*, "The Implementation of Knowledge Management Systems in Assets Reconciliation," in *The 5th International Conference on Information Technology and Engineering Application (ICIBA2016)*, Bina Darma University, Palembang, 2016, pp. 141-147.

[15] D. Meilia, *et al.*, "Evaluasi Infrastruktur Knowledge Sharing Pegawai pada Dinas Perpustakaan Provinsi Sumatera Selatan," in *Seminar Hasil Penelitian Sistem Informasi dan Teknik Informatika Ke-3 (SHaP-SITI2017)*, Palembang, 2017.

[16] B. Cohen, "Urbanization in developing countries: Current trends, future projections, and key challenges for sustainability," *Technology in Society,* vol. 28, pp. 63-80, 2006.https://doi.org/10.1016/j.techsoc.2005.10.005.

[17] S. Kemp, "Digital in 2018: World's internet users pass the 4 billion mark," in *We are social*, ed. Singapore: We Are Social Pte Ltd., 2018. Avalable: https://wearesocial.com/sg/blog/2018/01/global-digital-report-2018

[18] N. B. Grimm, *et al.*, "Global change and the ecology of cities," *Science,* vol. 319, pp. 756-760, 2008.https://doi.org/10.1126/science.1150195.

[19] L. A. Abdillah, "Indonesian's presidential social media campaigns," in *Seminar Nasional Sistem Informasi Indonesia (SESINDO2014)*, ITS, Surabaya, 2014. https://doi.org/10.13140/2.1.2549.3446

[20] E. M. Truman, "The G-20 and International Financial Institution Governance," Peterson Institute for International Economics, 2010.

[21] WorldBank, "Gross domestic product ranking table based on purchasing power parity (PPP)," World Bank2017.

[22] Statista. (2016). *Number of motor vehicles sold in the ASEAN region 2016, by country* Available: https://www.statista.com/statistics/583382/asean-motor-vehicle-sales-by-country/

[23] P. F. Belgiawan, *et al.*, "Understanding car ownership motivations among Indonesian students," *International Journal of Sustainable Transportation,* vol. 10, pp. 295-307, 2016.https://doi.org/10.1080/15568318.2014.921846.

[24] J. Jones and D. Hunter, "Consensus methods for medical and health services research," *BMJ: British Medical Journal,* vol. 311, p. 376, 1995.https://doi.org/10.1136/bmj.311.7001.376.

[25] J.-k. Kim, "The Effects of Teaching Image, Communal Sense, and Class Environment on Academic Procrastination in a University E-Learning Setting," *International Information Institute (Tokyo). Information*, vol. 20, p. 55, 2017.

[26] L. A. Abdillah, "Social media as political party campaign in Indonesia," *Jurnal Ilmiah MATRIK,* vol. 16, pp. 1-10, 2014.

[27] L. A. Abdillah, "Students learning center strategy based on e-learning and blogs," in *Seminar Nasional Sains dan Teknologi (SNST) ke-4 Tahun 2013*, Fakultas Teknik Universitas Wahid Hasyim Semarang 2013, pp. F.3.15-20. https://doi.org/10.13140/2.1.3466.8488.

[28] L. A. Abdillah, "Enriching Information Technology Course Materials by Using Youtube," in *The 5th International Conference On Artificial Intelligence, Computer Science and Information Technology (AICSIT2017)*, Bayview Beach Resort, Batu Ferringhi, Pulau Pinang, Malaysia, 2017, pp. 75-82.

[29] J.-S. Jeong*, et al.*, "A content oriented smart education system based on cloud computing," *International Journal of Multimedia and Ubiquitous Engineering,* vol. 8, pp. 313-328, 2013.https://doi.org/10.14257/ijmue.2013.8.6.31.

[30] M. Mohamed and A. A. Bahari, "Esl Instructors and Learners'views on The Use of Dropbox as A Sharing Tool in Writing Lessons," *Journal of Nusantara Studies (JONUS),* vol. 1, pp. 1-10, 2016.